\documentclass[aps,showpacs,floatfix,superscriptaddress,a4paper]{revtex4}
\usepackage{graphicx,amsmath,bbm,mathrsfs,amssymb,psfrag}
\newtheorem{definition}{Definition}
\newtheorem{remark}{Remark}

\newcommand{\CI}{{\mathbb{C}}}

\newcommand{\RI}{{\mathbb{R}}}

\newcommand{\cW}{{\mathcal{W}}}

\begin{document}

\author{Fabio Benatti}
\affiliation{Department of Physics, University of Trieste $\&$ INFN, Sezione di Trieste,\\
Strada Costiera 11, I-34051-Trieste,Italy\\
Email: benatti@ts.infn.it}

\author{Laure Gouba}
\affiliation{
The Abdus Salam International Centre for
Theoretical Physics (ICTP),\\
 Strada Costiera 11,
I-34151 Trieste Italy \\
Email: lgouba@ictp.it}

\title{Semi-Classical Localisation Properties of Quantum Oscillators on a Noncommutative Configuration Space }

\begin{abstract}
When dealing with the classical limit of two quantum mechanical oscillators on a noncommutative configuration space, 
the limits corresponding to the removal of configuration-space non\-commuta\-ti\-vi\-ty and position-momentum noncommutativity do not commute.     
We address this behaviour from the point of view of the phase-space localisation properties 
of the Wigner functions of coherent states under the two limits.
\end{abstract}

\maketitle

\section{Introduction}
The classical limit of quantum mechanics on a noncommutative configuration 
space has been recently studied \cite{fabio} by considering a system of two quantum harmonic oscillators whose spatial coordinates are 
themselves noncommuting operators with noncommutative parameter $\theta$ \cite{laure}.
One can then reduce to a system of classical harmonic oscillators on the commutative configuration space $\RI^2$ in two different ways: 
either removing the noncommutativity of the configuration space by letting $\theta\to 0$ and then quantumness, 
by letting $\hbar\to 0$ or, inverting the two limits.
Using the so-called anti-Wick quantization \cite{monique}, it has been shown in \cite{fabio}
\begin{enumerate}
\item that the 
two procedures do not commute and 
\item that, if one considers 
the simple free quadratic dynamics of the two noncommutative quantum oscillators, 
the asymmetry of the two limits is even stronger.
\end{enumerate}
Namely, letting $\theta\to 0$ first one regains 
the standard quantum mechanics of two independent harmonic oscillators; however, 
if $\hbar\to0$ with $\theta\neq 0$, no dynamics survives over the noncommutative configuration space.

Instead of considering the analogies with a quantum system in a magnetic 
field and a possible dynami\-cal explanation of the dimensional reduction \cite{delduc}, 
in the present paper, we are interested in the interpretation of such a 
nonexchangeability of the two limits by looking at the phase-space localisation properties
of the coherent states of the noncommutative quantum oscillators.
Indeed, in the standard classical limit, coherent states are common and powerful tools 
most to study semi-classical behaviours since their
Wigner functions are the closer to a Dirac delta in phase-space, the closer is $\hbar$ to $0$ \cite{ripamonti,stig}.

We shall show that, when $\hbar \rightarrow 0$ first, the Wigner function 
for the two noncommutative quantum oscillators vanishes as 
a pseudo distribution function over the classical $4$-dimensional phase-space, unless one can 
perform an integration over both position-like 
coordinates. This yields a well-defined marginal pseudo distribution with good 
localisation properties on a $2$-dimensional phase-space.
Furthermore, when the noncommutative quantum oscillators free dynamics is also accounted for, 
the elimination of the position like coordinates is not sufficient to obtain 
a well defined theory on the $2$-dimensional phase-space consisting of momentum-like coordinates. 
Indeed, the dynamical mixing of the first coordinate and its conjugate momentum requires a further 
integration which reduces the phase-space to a $1$-dimensional one, practically eliminating all memory 
of the initial system and its dynamics.

In Sect. \ref{sec3}, we summarize the main results obtained in \cite{fabio}, then 
in Sect. \ref{sec4} we calculate the Wigner functions of the coherent states 
and discuss their limits when $\hbar \rightarrow 0$ and $\theta \neq 0$. 
Concluding remarks are given in Sect. \ref{sec5}.

\section{The Classical limits of the Noncommutative Harmonic Oscillator}\label{sec3}

We consider the formalism of noncommutative quantum mechanics shortly reviewed in \cite{fabio} 
(for more details, see \cite{laure}) and study a model consisting of two noninteracting 
noncommutative quantum oscillators evolving according to the Hamiltonian operator
\begin{eqnarray}
\label{e7}
 \hat H = \sum_{i =1}^2\left(\frac{1}{2m}\hat P_i^2 +
\frac{1}{2}m\omega^2\hat X_i^2 \right)\ ,
\end{eqnarray}
where 
\begin{eqnarray}
\label{NCHA}
 \left[\hat X_i,\:\hat P_j\right] = i\hbar\delta_{i,j},\quad
 \left[\hat X_i,\:\hat X_j\right] = i\theta\epsilon_{i,j},\quad
 \left[\hat P_i, \hat P_j\right] = 0\ .
 \end{eqnarray}
One can associate to position and momentum operators creation and annihilation-like operators
$\hat A_{i}\,,\,\hat A^\dagger_{i}$, $i=1,2$ that satisfy the algebra
\begin{equation}\label{aacc}
\left[\hat A_{i},\: \hat A_{j}^\dagger \right]
= \delta_{ij},\quad
\left[ \hat A_{i}, \: \hat A_{j} \right] = 0 \ .
\end{equation}
The explicit expressions of the $\hat A^\#_i$ are as follows 
\begin{eqnarray}
\label{ca}
\hskip-.5cm
\hat A_{1} &=& \frac{1}{\sqrt{K^{\hbar,\theta}_+}}
\left( -\frac{\lambda^{\hbar,\theta}_+}{\hbar}\hat X_1 - i\hat P_1 -
i\frac{\lambda^{\hbar,\theta}_+}{\hbar}\hat X_2 +\hat P_2 \right),\\\label{ca1}
 \hat A_{1}^\dagger &=& \frac{1}{\sqrt{K^{\hbar,\theta}_+}}
\left( -\frac{\lambda^{\hbar,\theta}_+}{\hbar}\hat X_1 + i\hat P_1
+i\frac{\lambda^{\hbar,\theta}_+}{\hbar}\hat X_2 +\hat P_2 \right),\\
\label{caa}
\hskip -.5cm
\hat A_{2} &=& \frac{1}{\sqrt{K^{\hbar,\theta}_-}}
\left( \frac{\lambda^{\hbar,\theta}_-}{\hbar}\hat X_1 + i\hat P_1
-i\frac{\lambda^{\hbar,\theta}_-}{\hbar}\hat X_2 +\hat P_2 \right),\\\label{caa2}
\hat A_{2}^\dagger &=& \frac{1}{\sqrt{K^{\hbar,\theta}_-}}
\left( \frac{\lambda^{\hbar,\theta}_-}{\hbar}\hat X_1 - i\hat P_1 +
i\frac{\lambda^{\hbar,\theta}_-}{\hbar}\hat X_2 +\hat P_2 \right)\ ,
\end{eqnarray}
where
\begin{eqnarray}
\label{lambdas}
 \lambda^{\hbar,\theta}_\pm = \frac{1}{2}
\left( m\omega\sqrt{4\hbar^2 + m^2\omega^2\theta^2} \pm m^2\omega^2\theta\right),\quad
K^{\hbar,\theta}_\pm =& \lambda^{\hbar,\theta}_\pm\left(4 \pm \frac{2\lambda^{\hbar,\theta}_\pm\theta}{\hbar^2} \right).
\end{eqnarray}
Interestingly, the operators $\hat A^\#_j$ can be interpreted as proper
annihilation and creation operators as there is a vector in $\mathcal{H}_q$, namely
a Hilbert-Schmidt operator~\cite{laure}
\begin{equation}
\label{NNground}
\psi_{0}(\hat x_1, \hat x_2)
 = \exp{\Big(\frac{\beta^{\hbar,\theta}}{2\theta}(\hat x_1^2 +\hat x_2^2)\Big)}\ ,\quad \beta^{\hbar,\theta} =
 \ln( 1-\frac{\theta}{\hbar^2}\lambda^{\hbar,\theta}_-)
= -\ln(1 +\frac{\theta}{\hbar^2}\lambda^{\hbar,\theta}_+)\ .
\end{equation}
When interpreted as a Hilbert space vector it is such that
\begin{eqnarray}
\hat A_{1}\vert\psi_{0}\rangle = \hat A_2\vert\psi_0\rangle=0 \ ,
\end{eqnarray}
and corresponds to the normalized vacuum  \begin{equation}
\label{ground}
|0\rangle = \frac{\vert\psi_0\rangle}{\sqrt{\mathcal{N}}}\ , \quad
\mathcal{N} = \frac{\hbar^4}{2\hbar^2\lambda^{\hbar,\theta}_--\theta(\lambda^{\hbar,\theta}_-)^2}\ .
\end{equation}

In order to set up a proper framework for studying the classical and 
commutative limits, we introduce
the coordinate vector $r = (x_1,x_2,y_1,y_2)$ and the operator vector
$\hat r =\left(\hat X_1,\hat X_2, \hat P_1, \hat P_2\right)$.
Then, we construct the Weyl-like operators
\begin{equation}
\label{w1}
\hat W^{\hbar,\theta}(r) = \exp{\Big(\frac{i}{\mu^{\hbar,\theta}}
\left(r,\Omega\hat r\right)\Big)}\ ,
\end{equation}
where $\mu^{\hbar,\theta}$ is a parameter with the dimension of an action, 
and $\Omega$ is the symplectic matrix.

We shall focus upon two ways one can reach the classical, fully
commutative limit where  $\hbar=\theta=0$:
\begin{enumerate}
\item
by letting $\theta\to0$ first so to get to standard quantum mechanics
and then letting $\hbar\to0$;
\item
by letting $\hbar\to0$ first so to get to a generic noncommutative
system and then letting $\theta\to0$.
\end{enumerate}

In order to explore these two possibilities, we choose $\mu^{\hbar,\theta}$ such that
\begin{equation}\label{para}
\mu^{\hbar,0}:=\lim_{\theta \rightarrow 0}\mu^{\hbar,\theta} = \hbar; \quad
\mu^{0,\theta}:=\lim_{\hbar \rightarrow 0}\mu^{\hbar,\theta}=m\omega\theta\ .
\end{equation}
Indeed, the latter is the only natural constant with
the dimensions of an action when $\hbar=0$ in the model.
A natural choice is provided by (\ref{lambdas}):
\begin{equation}
\label{choice}
\mu^{\hbar,\theta}= \frac{\lambda^{\hbar,\theta}_+}{m\omega}\ ,
\end{equation}
a quantity fulfilling \eqref{para}, whereas $\lim_{\hbar\to 0}\lambda^{\hbar,\theta}_{-}=0$ as one 
deduces from the limit behaviours of \- the $\hbar$, $\theta$-dependent quantities reported in the Appendix.

One can thus rewrite the Weyl operators~(\ref{w1}) in the form
\begin{eqnarray}
\label{w2}
\hat W(z^{\hbar,\theta}_r) = \exp{\Big(z^{\hbar,\theta}_{1r}\hat A^\dag_1\,
+\,z^{\hbar,\theta}_{2r}\hat A^\dag_2\,-\,(z^{\hbar,\theta}_{1r})^*\hat A_1\,-\,(z^{\hbar,\theta}_{2r})^*\hat A_2\Big)}\ ,
\end{eqnarray}
with $z^{\hbar,\theta}_r=(z^{\hbar,\theta}_{1r},z^{\hbar,\theta}_{2r})$ a
two-dimensional complex vector whose real and imaginary parts are
connected to the real four-dimensional vector $r$ by
\begin{equation}
\label{XZ}
\begin{pmatrix}
\mathcal{R}e(z^{\hbar,\theta}_{1r})\cr
\mathcal{R}e(z^{\hbar,\theta}_{2r})\cr
\mathcal{I}m(z^{\hbar,\theta}_{1r})\cr
\mathcal{I}m(z^{\hbar,\theta}_{2r})
\end{pmatrix}= \hat{J}^{\hbar,\theta} r\ ,\quad \textrm{where}
\end{equation}
\begin{equation}
\hat{J}^{\hbar,\theta} = \frac{1}{2\mu^{\hbar,\theta}(\lambda^{\hbar,\theta}_{+} +\lambda^{\hbar,\theta}_{-})}
\begin{pmatrix}
 \lambda^{\hbar,\theta}_{-}\sqrt{K^{\hbar,\theta}_+} & 0 & 0 & -\hbar\sqrt{K^{\hbar,\theta}_+}\cr
 -\lambda^{\hbar,\theta}_+\sqrt{K^{\hbar,\theta}_{-}} & 0 & 0 & -\hbar\sqrt{K^{\hbar,\theta}_{-}}\cr
 0 & \lambda^{\hbar,\theta}_{-}\sqrt{K^{\hbar,\theta}_+} & \hbar\sqrt{K^{\hbar,\theta}_+} & 0\cr
 0 & \lambda^{\hbar,\theta}_{+}\sqrt{K^{\hbar,\theta}_{-}} & -\hbar\sqrt{K^{\hbar,\theta}_{-}} & 0
\end{pmatrix}\ .
\end{equation}
 
By using the ground state~(\ref{ground}) and the relations~(\ref{aacc}), in analogy with the coherent 
states of standard quantum mechanics, we now introduce the states
\begin{equation}
\label{coh}
 \vert z^{\hbar,\theta}_r\rangle = \hat W(z^{\hbar,\theta}_r)\vert 0\rangle\,
 =\,\exp{\Big(-\frac{\|z^{\hbar,\theta}_r\|^2}{2}\Big)}\, \exp{\Big(z^{\hbar,\theta}_{1r}\hat A_1^\dag+z^{\hbar,\theta}_{2r}
 \hat A^\dag_2\Big)}\,\vert0\rangle\ ,
\end{equation}
where $\|z^{\hbar,\theta}_r\|^2=|z^{\hbar,\theta}_{1r}|^2+|z^{\hbar,\theta}_{2r}|^2$.
Because of the algebraic relations~(\ref{aacc}), it follows that
\begin{equation}
\label{cohNC}
\hat A_1\,\vert z^{\hbar,\theta}_r\rangle\,=\,z^{\hbar,\theta}_{1r}\,\vert z^{\hbar,\theta}_r\rangle\
,\quad \hat A_2\vert z^{\hbar,\theta}_r\rangle\,=
\,z^{\hbar,\theta}_{2r}\,\vert z^{\hbar,\theta}_r\rangle\ .
\end{equation}
Despite the fact that they are not minimal indeterminacy states as standard coherent states, 
they have a Gaussian character and constitute an over-complete set \cite{fabio}:
\begin{equation}
\frac{1}{\pi^2}\int_{\CI^2} {\rm d}z^{\hbar,\theta}\, \vert z^{\hbar,\theta}\rangle\langle z^{\hbar,\theta}\vert
= \hat{1}\ .
\end{equation}

By means of these coherent states one can set up the so-called anti-Wick quantisation scheme 
which is based on specific quantization and de-quantization 
maps from a commutative $C^*$ algebra $\mathcal{A}_4$ with identity into 
the quantum Weyl $C^*$ algebra $\cW^{\hbar,\theta}$ generated by the Weyl operators.

One may take as $\mathcal{A}_4$ the $C^*$ algebra $C^\infty(\mathbb{R}^4)\cup 1$ of infinity differentiable functions 
over the $4$-dimensional phase-space which go 
to zero at infinity with all their derivatives to which the identity is added.
In this way, any $F\in\mathcal{A}_4$ is such that whenever one of its argument diverges, the limit exists and yields 
a function of the remaining arguments which does not  necessarily vanish. 
In particular, in the following, we shall consider cases where the arguments $x_{1,2}$ in $F(x_1,x_2,y_1,y_2)$ go to 
$\pm\infty$, so that 
\begin{equation}
\label{limitfunct}
\lim_{x_1,x_2\to\pm\infty}F(x_1,x_2,y_1,y_2)=F_\infty(y_1,y_2)\in \mathcal{A}_2=C^\infty(\mathbb{R}^2)\cup 1.
\end{equation}

\begin{definition}
Given the commutative $C^*$ algebra $\mathcal{A}_4$ and the Weyl algebra $\cW^{\hbar,\theta}$ be the $C^*$ 
algebra generated by the Weyl operators~(\ref{w1}), 
a function $F\in \mathcal{A}_4$ is turned into an operator element of the Weyl $C^*$ 
algebra by the positive unital quantisation map
$\gamma_{0\mapsto(\hbar,\theta)}:\mathcal{A}_4\mapsto \cW^{\hbar,\theta}$ defined by
\begin{equation}
\label{qNC}
\mathcal{A}_4\ni F\mapsto\gamma_{0\mapsto(\hbar,\theta)}[F]=:\hat F^{\hbar,\theta}\in\cW^{\hbar,\theta}\ ,\quad
\hat F^{\hbar,\theta}=\frac{J^{\hbar,\theta}}{\pi^2}\int_{\RI^4} {\rm d}r\, F(r)\,
\vert z^{\hbar,\theta}_r\rangle\langle z^{\hbar,\theta}_r\vert\ .
\end{equation}
Vice versa, any operator in $\cW^{\hbar,\theta}$ can be mapped into a function $F\in \mathcal{A}_4$ 
by the de-quantisation unital
map $\gamma_{(\hbar,\theta)\mapsto0}:\cW^{\hbar,\theta}\mapsto \mathcal{A}_4$ defined by:
\begin{equation}
\label{dqNC}
\cW^{\hbar,\theta}\ni\hat X\mapsto
\gamma_{(\hbar,\theta)\mapsto0}[\hat X]=:X^{\hbar,\theta}(r)\in \mathcal{A}_4\ ,
\quad X^{\hbar,\theta}(r)=\langle z^{\hbar,\theta}_r\vert\,\hat X\,\vert z^{\hbar,\theta}_r\rangle\ .
\end{equation}
\end{definition}
\medskip

The classical limit of the noncommutative quantum oscillators can then be
performed by means of the following map from $\mathcal{A}_4$ into itself:
\begin{equation}
\label{qdqNC}
\mathcal{A}_4\ni F\mapsto
F^{\hbar,\theta}=\gamma_{(\hbar,\theta)\mapsto0}\circ\gamma_{0\mapsto(\hbar,\theta)}[F]\in \mathcal{A}_4\ ,
\end{equation}
which is such that 
\begin{eqnarray}
\label{qdqNC1}
F^{\hbar,\theta}(r) &=& \frac{J^{\hbar,\theta}}{\pi^2}
\int_{\mathbb{R}^4} {\rm d}r'\, F(r')\,\left\vert \langle z^{\hbar,\theta}_r\vert z^{\hbar,\theta}_{r'}
\rangle\right\vert^2\\\label{qdqNC2}
&=&\frac{1}{\pi^2}
\int_{\mathbb{R}^4}{\rm d}w\,
e^{-\|w\|^2}\,F(r+h^{\hbar,\theta}(w)),
\end{eqnarray}

where $h^{\hbar,\theta}(w)=(f^{\hbar,\theta}(w_1,w_2)\ ,f^{\hbar,\theta}(w_3,w_4)\ ,g^{\hbar,
\theta}(w_3,w_4)\ ,-g^{\hbar,\theta}(w_1,w_2)\Big)$ with
\begin{eqnarray}
\label{f12}
&&
f^{\hbar,\theta}(w_1,w_2)=\frac{\mu^{\hbar,\theta}\sqrt[4]{4\hbar^2 + m^2\omega^2\theta^2}}{2\sqrt{m\omega}\hbar}
 \left(\frac{w_1}{\sqrt{\gamma^{\hbar,\theta}_+}}+\frac{w_2}{\sqrt{\gamma^{\hbar,\theta}_{-}}}\right)\\
\label{f12a}
&&
g^{\hbar,\theta}(w_1,w_2)=\frac{\mu^{\hbar,\theta}\sqrt{m\omega}}{\sqrt[4]{4\hbar^2 + m^2\omega^2
\theta^2}}\left(\frac{w_1}{\sqrt{\gamma^{\hbar,\theta}_+}}-
\frac{w_2}{\sqrt{\gamma^{\hbar,\theta}_{-}}}\right)\ ,\\
\label{g12}
&&
\gamma^{\hbar,\theta}_\pm = \frac{1}{2}\Big(1 \pm\frac{m\omega\theta}
{\sqrt{4\hbar^2 + 2m^2\omega^2\theta^2}}\Big)\ .
\end{eqnarray}
The behaviours of the above quantities reported in the Appendix and allow one to draw the following conclusions: 

\begin{itemize}
\item
When $\theta\to0$, the expressions \eqref{A13}-\eqref{A15} in the Appendix yield the limits
$$
\gamma^{\hbar,0}_\pm=\frac{1}{2}\ ,\quad f^{\hbar,0}(x,y)=\sqrt{\frac{\hbar}{m\omega}}(x+y)\ ,
\quad g^{\hbar,0}(x,y)=\sqrt{\hbar m\omega}(x-y)\ ,
$$
so that 
\begin{eqnarray}
\label{NCQ2Q2C}
\lim_{\hbar\rightarrow 0}\lim_{\theta\to0}F^{\hbar,\theta}(r)=F(r)\in\mathcal{A}_4\ .
\end{eqnarray}
\item
By letting $\hbar\to0$, from the expressions \eqref{A10}-\eqref{A12} in the Appendix, one sees that, while
the function $g^{\hbar,\theta}(x,y)$ converges to  
\begin{equation}
\label{fg}
g^{0,\theta}(x,y)=m\omega\sqrt{\theta}\left(\frac{x}{\sqrt{1+\frac{1}{\sqrt{2}}}}-\frac{y}{\sqrt{1-\frac{1}{\sqrt{2}}}}\right)\ ,
\end{equation}
the function $f^{\hbar,\theta}(x,y)$ diverges as $1/\hbar$. Therefore, according to the discussion before \eqref{limitfunct},
the integrated function $F(x_1,x_2,y_1,y_2)$ in \eqref{qdqNC2} 
becomes a function of $y_1$ and $y_2$, only.
Then, by letting $\theta\to 0$, one further removes the noncommutativity of 
the configuration space so that 
\begin{equation}
\label{NCQ2NCig2C}
\lim_{\theta\to0}\lim_{\hbar\to0}F^{\hbar,\theta}(r)= F_\infty(y_1,y_2)\in\mathcal{A}_2\ .
\end{equation}
\end{itemize}

Hence, the removal of quantum noncommutativity followed by the removal of 
configuration space noncommutativity do not get back to the initial 
commutative algebra of functions over $\RI^4$, $\mathcal{A}_4$, rather to the commutative algebra $\mathcal{A}_2$ 
of functions on $\mathbb{R}^2$. 
Therefore, the two de-quantizing limits do not commute:
\begin{equation}
\label{NCL}
\lim_{\theta\to0}\lim_{\hbar\to0}F^{\hbar,\theta}(r)\,\neq\,\lim_{\hbar\to0}\lim_{\theta\to0}F^{\hbar,\theta}(r)\ .
\end{equation}

Let us now consider the time-evolution generated by the Hamiltonian~(\ref{e7}), 
using as dimensional action, not $\hbar$,
but the parameter$\mu^{\hbar,\theta}$ in~(\ref{choice}).
The unitary time-evolution on the noncommutative Hilbert space $\mathcal{H}_q$ is thus given by
\begin{equation}
\label{time-evolutor}
\hat U_t = \exp{\Big(-\frac{it}{\mu^{\hbar,\theta}}\hat H\Big)}\ .
\end{equation}
Its action on the Weyl operators in the forms~(\ref{w1}) and~(\ref{w2}) is easily computed to be
\begin{equation}
\label{time-evolutor1}
\hat U^\dag_t\,\hat W^{\hbar,\theta}(r)\,\hat U_t = \hat W^{\hbar,\theta} (r_{-t})
\end{equation}
where
\begin{equation}
\label{Weylt1}
r_{-t} = A^{\hbar,\theta}_{-t}\, r\ , \quad
 A^{\hbar,\theta}_{-t} = \left(
 \begin{array}{cccc}
  \cos\omega^{\hbar,\theta}_+t & 0 & \sin\omega^{\hbar,\theta}_+t & 0\\
  0 & \cos\omega^{\hbar,\theta}_{-}t & 0 & \sin\omega^{\hbar,\theta}_{-}t\\
  -\sin\omega^{\hbar,\theta}_+t & 0 & \cos\omega^{\hbar,\theta}_+t & 0\\
  0 & -\sin\omega^{\hbar,\theta}_{-}t & 0& \cos\omega^{\hbar,\theta}_{-}t
 \end{array}
 \right)\ ,
\end{equation}
with the oscillation frequencies given by
\begin{equation}
\label{frequencies}
\omega^{\hbar,\theta}_\pm =
\frac{\lambda^{\hbar,\theta}_\pm}{m\mu^{\hbar,\theta}}
\end{equation}
The various limit behaviours of these quantities and of the time-evolution matrix $A^{\hbar,\theta}_t$ 
are given in the Appendix.

Since the ground state $\vert 0\rangle$ in~(\ref{ground}) is left invariant by $U_t$, 
one finds that the time-evolution of the quantised function
in~(\ref{qNC}) is given by
\begin{equation}
\label{qNCt}
\cW^{\hbar,\theta}\ni\hat F^{\hbar,\theta}_t=\hat U_t^\dag\,\hat F^{\hbar,\theta}\, \hat U_t=
\frac{J^{\hbar,\theta}}{\pi^2}\int_{\RI^4} {\rm d}r\, F^{\hbar,\theta}_t(r)\,
\vert z^{\hbar,\theta}_r\rangle\langle z^{\hbar,\theta}_r\vert\ ,
\end{equation}
where it has been used that $\hbox{Det}(A^{\hbar,\theta}_{t})=1$ and  
$F^{\hbar,\theta}_t(r)=F(A^{\hbar,\theta}_t\,r)$ has been set.
Then,~(\ref{qdqNC}) yields
\begin{eqnarray}
\nonumber
F^{\hbar,\theta}_t(r)&=&\gamma_{(\hbar,\theta)\mapsto0}
\Big[\hat U_t^\dag\gamma_{0\mapsto(\hbar,\theta)}[F]\hat U_t\Big](r)\\\label{qdqNCt}
&=&\frac{1}{\pi^2}
\int_{\mathbb{R}^4}{\rm d}w\,
e^{-\|w\|^2}\,F^{\hbar,\theta}_t\Big(r+h^{\hbar,\theta}(w)\Big),
\end{eqnarray}
where $h^{\hbar,\theta}(w)=\Big(f^{\hbar,\theta}(w_1,w_2)\ ,f^{\hbar,\theta}(w_3,w_4)\ ,g^{\hbar,\theta}(w_3,w_4)\ ,
- g^{\hbar,\theta}(w_1,w_2)\Big)$,
with the functions $f^{\hbar,\theta}, g^{\hbar,\theta}$ as in~(\ref{f12}), (\ref{f12a}).

\begin{itemize}
\item
By letting first $\theta\to0$ in~(\ref{qdqNCt}) one recovers the commutative quantum mechanical context;
indeed from~(\ref{frequencies}) one has
$\lim_{\theta\rightarrow 0}\omega_\pm^{\hbar,\theta} = \omega $, whence
\begin{equation}
F^\hbar_t(r)= \lim_{\theta\rightarrow 0}F^{\hbar,\theta}_t(r) =
 \frac{1}{\pi^2}
 \int_{\mathbb{R}^4}dw\, e^{-||w||^2}\, F\Big( A^{\hbar,0}_{-t}\,(r + h^{\hbar,0}(w))\Big) \ ,
\end{equation}
where $A^{\hbar,0}_{t}$ is given in \eqref{A21} of the Appendix and coincides with the
classical time-evolution matrix $A_t=A^{0,0}_t$. 
Thence, in the classical limit $\hbar\to0$, one obviously recovers the time-evolution 
of two classical harmonic oscillators:
\begin{equation}
\label{clt}
\lim_{\hbar\rightarrow 0}F^\hbar_t(r)= F( A_{-t}\, r)\ .
\end{equation}
\item
By letting $\hbar\to0$ in~(\ref{qdqNCt}), one would expect to obtain a dynamical system over the noncommutative 
configuration space context. Using the time-evolution matrix $A^{0,\theta}_t=\lim_{\hbar\to0}A^{\hbar,\theta}_t$ given
in \eqref{A24} of the Appendix, one finds
\begin{eqnarray}
\nonumber
F^\theta_t(y_2)=\lim_{\hbar\rightarrow 0}F^{\hbar,\theta}_t(r)
&=& \frac{1}{\pi^2}\int_{\mathbb{R}^4}{\rm d}w\,
 e^{-\|w\|^2}\\\nonumber
 &\times& F_\infty\left(y_2 + 2m\omega\sqrt{\theta}
 \left(\frac{w_2}{\sqrt{1-\frac{1}{\sqrt{2}}}} -
 \frac{w_1}{\sqrt{1+\frac{1}{\sqrt{2}}}}\right)\right)\\
\label{NCQ2NCat}
&=& \frac{1}{\sqrt{\pi}}\int_{\mathbb{R}}dv_2\;
  e^{-v_2^2}F_\infty(y_2 +2m\omega\sqrt{\theta}v_2)\ ,
\end{eqnarray}
where the function $F_\infty(y_2)$ denotes the limit
$\displaystyle \lim_{x_1,x_2,y_1\to+\infty}F(r)$ and is effectively 
a function of $y_2$. It follows that there is no dynamics on the noncommutative 
configuration space. Furthermore, the full classical limit yields
\begin{equation}
\label{nodyn}
 \lim_{\theta\to 0}F_{t}^{\theta}(y_2)= F_\infty({y_2})\ .
\end{equation}
\end{itemize}

Therefore, starting with the continuous functions over $\RI^4$, 
letting the dynamics act and then removing the standard noncommutativity before 
removing the configuration space noncommutativity one loses track of the time-evolution and even reduces, 
after the complete classical limit, the domain of definition of the continuous functions from $\RI^4$ to $\RI$.

\section{Phase-space interpretation of the non-exchangeability of the limits}
\label{sec4}

In standard quantum mechanics, coherent states are most useful tools
to study the classical limit; indeed, these states have very good localisation properties in
phase-space. This fact can be best appreciated by looking at their Wigner function, namely at the pseudo probability 
distribution on phase space associated with any quantum state. 
For sake of simplicity, consider a system with one degree of freedom described by position and momentum operators 
$\hat{q}$ and $\hat{p}$ such that $[\hat{q},\hat{p}]=i\hbar$, or by annihilation and creation operators 
$$
\displaystyle\hat{a}=\frac{\hat{q}+i\hat{p}}{\sqrt{2\hbar}}\ ,\quad \displaystyle\hat{a}^\dag=\frac{\hat{q}-i\hat{p}}{\sqrt{2\hbar}}
$$ 
satisfying $[\hat{a}\,,\,\hat{a}^\dag]=1$. In the above we consider suitably rescaled position and momentum operators 
so that $\hbar$ is an a-dimensional parameter. 

Given the phase-space point $r_0=(q_0,p_0)$, the associated coherent state is given by acting on the vacuum state 
$\vert 0\rangle$ with the Weyl operator $\displaystyle\hat W^\hbar(r_0)={\rm e}^{i/\hbar(p_0\hat{q}-q_0\hat{p})}$ :
$$
\vert z_{r_0}\rangle={\rm e}^{z_{r_0}\hat{a}^\dag-z_{r_0}^*\hat{a}}\vert 0\rangle
=W^\hbar(r_0)\vert 0\rangle\ ,\quad z_{r_0}=\frac{q_0+ip_0}{\sqrt{2\hbar}}\ .
$$ 
Its Wigner function $R^\hbar_{q_0,p_0}(q,p)$ is then defined as the Fourier transform of the characteristic function 
$$
C_{z_{r_0}}(x,y)=\langle z_{r_0}\vert {\rm e}^{i/\hbar(y\hat{q}-x\hat{p})}\vert z_{r_0}\rangle={\rm e}^{-\|r_0\|^2/2}\,
{\rm e}^{i/\hbar(yq_0-xp_0)}\ .
$$
Namely, 
\begin{equation}
\label{wig0}
R^\hbar_{q_0,p_0}(q,p):=\int \frac{dx\,dy}{2\pi\hbar}\,{\rm e}^{i/\hbar(xp-yq)}\,C_{z_{r_0}}(x,y)=
\frac{1}{\pi\hbar}\,{\rm e}^{-\left((q-q_0)^2 + (p-p_0)^2\right)/\hbar}\ .
\end{equation}
One thus sees that in the limit $\hbar\rightarrow 0$, the Wigner function becomes a Dirac delta 
at $(q_0,p_0)$. 
Furthermore, given a reasonably smooth classical Hamiltonian $H(q,p)$ and the associated phase-space trajectory\-\break
$r_0=(q_0,p_0)\mapsto r_t=(q_t,p_t)$, the corresponding quantum dynamics
$\hat{U}_t=\exp(-it\hat{H}/\hbar)$ maps coherent states $\vert z_{r_0}\rangle$ into states $\hat{U}_t\vert z_{r_0}\rangle$ 
whose Wigner function, in the limit $\hbar\to0$, is localised around the classical trajectory $r_0\mapsto r_t$ \cite{monique}. 

\begin{remark}
\label{remWigner}
The Wigner function 
$$
R^\hbar_\rho(q,p):=\int \frac{dx\,dy}{2\pi\hbar}\,{\rm e}^{i/\hbar(xp-yq)}\,{\rm Tr}\Big(\rho\,\hat W^\hbar(r)\Big) ,\quad r=(x,y)\ ,
$$ 
of a generic quantum state $\rho$ behaves as a pseudo probability distribution in phase-space; indeed, 
though it is not in general positive definite, it is normalised 
$\int dq\,dp\,R^\hbar_\rho(q,p)=1$ and allows to compute the mean values of quantum observables as phase-space 
integrals with respect to $R^\hbar_\rho(q,p)$:
$$
{\rm Tr}(\rho\,\hat A)=\int \frac{dq\,dp}{2\pi\hbar}\, R^\hbar_\rho(q,p)\, A^\hbar(q,p)\ ,
$$
where the phase-space function $A^\hbar(q,p)$ associated with the Hilbert space operator $\hat A$ is given by
$$
A^\hbar(q,p)=\int \frac{dx\,dy}{2\pi\hbar}\,{\rm e}^{-i/\hbar(xp-yq)}\, {\rm Tr}\Big(\hat{A}\,\hat W^\hbar(-r) \Big)\ ,\quad r=(x,y)\ .
$$
\end{remark}

Using creation and annihilation operators, 
one could express the Wigner pseudo distribution as a function of $z\in \mathbb{C}$ as follows
\begin{equation}
\label{wig1a}
S_{z_{r_0}}(z):=\int \frac{d^2u}{\pi}\,{\rm e}^{2i\mathcal{I}_m(u^*z)}\,
\langle z_{r_0}\vert{\rm e}^{u\hat{a}^\dag-u^*\hat{a}}\vert z_{r_0}\rangle
=2{\rm e}^{-2|z-z_{r_0}|^2}\ ,
\end{equation}
where $d^2u=d\mathcal{R}_e(u)\,d\mathcal{I}_m(u)$. 

\subsection{Noncommutative quantum harmonic oscillator: localisation properties}\label{ssec1}

Since the structure and properties of the Wigner function $S_{r_{z_0}}(z)$ only depend on the algebraic canonical 
commutation relations, we can extend it to cover the system of two noncommutative quantum harmonic oscillators.
Given a fixed phase point $r_0=(x_{10},x_{20},y_{10},y_{20})$, with position-like coordinates $(x_{10},x_{20})$ and 
momentum-like coordinates $(y_{10},y_{20})$, we associate to it the coherent state  
$\vert z_{r_0} \rangle^{\hbar,\theta} = \hat W^{\hbar,\theta}(z_{r_0})\vert 0\rangle$ given in \eqref{coh} with 
$z_{r_0}=\hat Jr_0$ as in \eqref{XZ}. Then, we define its Wigner function in the complex representation as
\begin{eqnarray}\nonumber
S_{z_{r_0}}(z)&=&\int \frac{d^2u}{\pi}\frac{d^2v}{\pi}
{\rm e}^{2i\mathcal{I}m(u^*z_1 + v^*z_2)}\,\langle z_{r_0}\vert
{\rm e}^{u\hat{A}_1^\dag+v\hat{A}_2^\dag-u^*\hat{A}_1-v^*\hat{A}_2}\vert z_{r_0}\rangle\\\label{wig1}
&=& 4e^{-2\|z-z_{r_0}\|^2}\ .
\end{eqnarray}
where $z=(z_1,z_2)\in\mathbb{C}^2$.
Its structure as a phase-space function (see \eqref{wig0}) shows up by substituting $z=\hat{J}r$, thus connecting complex 
vectors in $\mathbb{C}^2$ with phase points $r=(x_1,x_2,y_1,y_2)$ in $\mathbb{R}^4$ (compare \eqref{XZ}).
Then, using \eqref{XZ} we get the following distribution on $\mathbb{R}^4$ 
\begin{equation}
R^{\hbar,\theta}_{r_0}(r)= \frac{J^{\hbar,\theta}}{\pi^2}\,{\rm e}^{-2\|\hat{J}^{\hbar,\theta}(r-r_0)\|^2}
\label{wig3}
\end{equation}
where the determinant $J^{\hbar,\theta}=\hbox{Det}\left(\hat{J}^{\hbar,\theta}\right)$ accounts for the 
normalisation upon integration of $R_{r_0}(r)$ over $\mathbb{R}^4$.
In the limit $\theta\rightarrow 0$, using the expression of $\hat{J}^{\hbar,0}:=\lim_{\theta\to0}\hat{J}^{\hbar,\theta}$ in 
equation \eqref{A11} of the Appendix, one gets
\begin{eqnarray}\nonumber
R^\hbar_{r_0}(r) &:=& \lim_{\theta\rightarrow 0} R^{\hbar,\theta}_{r_0}(r)\\\label{wig4}
&=& \frac{1}{(\pi\hbar)^2}
 e^{-\frac{m\omega}{\hbar}\left((x_1-x_{10})^2 + (x_2- x_{20})^2\right)-\frac{1}{\hbar m\omega}\left((y_1-y_{10})^2 +(y_2-y_{20})^2\right)}\ .
\end{eqnarray}
According to \eqref{wig0}, the latter is the Wigner function of the two-oscillator coherent state obtained by acting
on the vacuum state with the Weyl operator
$$
W^\hbar(r_0)={\rm e}^{i/\hbar\left(y_{10}\hat{x}_1+y_{20}\hat{x}_2-x_{10}\hat{y}_1-x_{20}\hat{y}_2\right)}\ ,
$$
where the vacuum state is annihilated by
\begin{equation}
\label{creatosc}
\hat{a}_1=\sqrt{\frac{m\omega}{2\hbar}}\hat{x}_1\,+\,\frac{i}{\sqrt{2m\omega\hbar}}\hat{y}_1\ ,\quad
\hat{a}_2=\sqrt{\frac{m\omega}{2\hbar}}\hat{x}_2\,+\,\frac{i}{\sqrt{2m\omega\hbar}}\hat{y}_2\ .
\end{equation}
The Dirac delta localisation properties of $R^\hbar_{r_0}(r)$ on the phase-space 
$\mathbb{R}^4$ (when $\hbar\rightarrow 0$) are typical of such a system and corresponds to its becoming a 
system of two classical oscillators in the standard classical limit. 

On the contrary, from \eqref{A11} in the Appendix, one sees that, in the limit $\hbar \rightarrow 0$, the Jacobian 
$J^{0,\theta}:=\lim_{\hbar\to0}J^{\hbar,\theta}=0$,  while  
\begin{equation}
\label{wig5}
{\rm e}^{-2\|\hat{J}^{\hbar,\theta}(r-r_0)\|^2}\longmapsto{\rm e}^{-2\|\hat{J}^{0,\theta}(r-r_0)\|^2}={\rm e}^{-\frac{1}{{\theta m^2\omega^2}}
\left((y_1-y_{10})^2 + (y_2-y_{20})^2\right)}\ .
\end{equation}
As a consequence, when one computes mean values of integrable phase-space functions $F(x_1,x_2,y_1,y_2)$
with respect to the pseudo distribution $R^{\hbar,\theta}_{r_0}(r)$ and takes
the limit $\hbar \rightarrow 0$, the result always vanishes unless the integrated functions $F(x_1,x_2,y_1,y_2)$ 
do not depend on $(x_1,x_2)$. 
Then, by removal of quantum $(q,p)$ noncommutativity, a meaningful theory on noncommutative configuration-space 
is only possible if one-considers mean values of observables that correspond to functions of the form
$F(y_1,y_2)$. Indeed, their mean values  are calculated by means of the reduced marginal distribution that results from 
integrating $R^{\hbar,\theta}_{r_0}(r)$ over $x_1,x_2$. This yields
\begin{eqnarray}\nonumber
R^{\hbar,\theta}_{y_{10},y_{20}}(y_1,y_2) &:=& \int dx_1dx_2 R^{\hbar,\theta}_{r_0}(x_1,x_2,y_1,y_2)\\\nonumber
&=& \frac{J^{\hbar,\theta}}{\pi({(a^{\hbar,\theta})}^2+{(b^{\hbar,\theta})}^2)}\,
{\rm e}^{-\frac{J^{\hbar,\theta}}{({a^{\hbar,\theta})}^2+{(b^{\hbar,\theta})}^2}
\left((y_1-y_{10})^2 + (y_2-y_{20})^2\right)}\\\label{wig6}
\end{eqnarray}
where
\begin{equation}
\label{wig7}
a^{\hbar,\theta}
=\frac{\lambda_-^{\hbar,\theta}\sqrt{K^{\hbar,\theta}_+}}{2\mu^{\hbar,\theta}\left(\lambda_+^{\hbar,\theta}+\lambda_-^{\hbar,\theta}\right)}
\ ,\qquad 
b^{\hbar,\theta}
=\frac{\lambda_+^{\hbar,\theta}\sqrt{K^{\hbar,\theta}_-}}{2\mu^{\hbar,\theta}\left(\lambda_+^{\hbar,\theta}+\lambda_-^{\hbar,\theta}\right)}
\ .
\end{equation}
Using the limiting behaviours in \eqref{A5}-\eqref{A10} in the Appendix, one gets
$$
\lim_{\hbar\rightarrow 0}\frac{J^{\hbar,\theta}}{{(a^{\hbar,\theta})}^2 +{(b^{\hbar,\theta})}^2} = \frac{1}{m^2\omega^2\theta}\ ,
$$
whence, with $y=(y_1,y_2)$ and $y_0=(y_{10},y_{20})$ in $\mathbb{R}^2$,
\begin{equation}
\label{wig8}
\lim_{\hbar\rightarrow 0}\frac{J^{\hbar,\theta}\ R^{\hbar,\theta}_{y_{10},y_{20}}(y_1,y_2)}{{(a^{\hbar,\theta})}^2 + {(b^{\hbar,\theta})}^2}= 
\frac{1}{\pi m^2\omega^2\theta}\, {\rm e}^{-\frac{1}{m^2\omega^2\theta}\left((y_1 -y_{10})^2 + (y_2-y_{20})^2\right)}=:
R^\theta_{y_0}(y) \ .
\end{equation}
By comparison with the standard expression \eqref{wig0}, $R^\theta_{y_0}(y)$ is the Wigner function 
of a system with one degree of freedom associated with a coherent state $\vert z^\theta_{y_0}\rangle$, generated by a Weyl operator of the form 
$$
\hat{W}^\theta(y_0)={\rm e}^{i(y_{20}\hat{x}_1+y_{10}\hat{x}_2)/m\omega\theta}\ ,
$$
when acting on a vacuum state associated to the one-degree of freedom annihilation operator
\begin{equation}
\label{creat}
\hat{b}=\frac{\hat{x}_1\,+\,i\,\hat{x}_2}{\sqrt{2\theta}}\ .
\end{equation}
Such a Wigner function possesses Dirac delta localisation properties around $(y_{10},y_{20})$ when, by letting $\theta \rightarrow 0$, 
one removes the noncommutativity of the configuration-space. 

Therefore, when $\hbar\rightarrow 0$, a meaningful kinematics can emerge on the noncommutative configuration space,
only by reducing from a four dimensional phase-space to a two dimensional one; the latter consists of  momentum-like 
coordinates corresponding to the non commuting positions $\hat{x}_{1,2}$.

\subsection{Free non-commutative oscillator dynamics: localisation properties}\label{ssec2}

Under the time-evolution operator $\hat U(t)$ in \eqref{time-evolutor}, using \eqref{time-evolutor1} and the fact that the 
vacuum state is left invariant by $\hat{U}^{\hbar\,\theta}_t$, coherent states evolve in time into coherent states:
\begin{equation}
\label{Weylt0}
\vert z^{\hbar,\theta}_{r_0}\rangle\longmapsto U^{\hbar,\theta}_t\vert z^{\hbar,\theta}_{r_0}\rangle
=\vert z^{\hbar,\theta}_{r^{\hbar,\theta}_{-t}(r_0)}\rangle\ ,
\end{equation}
where, according to \eqref{Weylt1},
\begin{equation}
\label{tXZ}
z^{\hbar,\theta}_{r^{\hbar,\theta}_{-t}(r_0)}=\hat{J}^{\hbar,\theta}A^{\hbar,\theta}_{-t}\ r_0\ .
\end{equation}
Using \eqref{wig3}, the corresponding Wigner function evolves in time as follows:
\begin{equation}
\label{wigt}
R^{\hbar,\theta}_{r_0}(r)\longmapsto R^{\hbar,\theta}_{r^{\hbar,\theta}_t(r_0)}(r)=\frac{J^{\hbar,\theta}}{\pi^2}\,
{\rm e}^{-2\left\|\hat{J}^{\hbar,\theta}\left(r - r^{\hbar,\theta}_{-t}(r_0)\right)\right\|^2}\ .
\end{equation}

Because of \eqref{A11} and \eqref{A21} in the Appendix, letting $\theta\to0$, from 
$R^{\hbar,\theta}_{r^{\hbar,\theta}_t(r_0)}(r)$ one recovers the Wigner function of two free harmonic oscillator which is 
localised around the classical trajectory $r_0\mapsto A_{-t}r_0$. 

Instead, when letting $\hbar\to0$ first, as explained in the previous section, one has to start at time $t=0$ with the reduced Wigner function 
$R^{\hbar,\theta}_{y_0}(y)$ in \eqref{wig4} with $y_0=(y_{10},y_{20})$ and $y=(y_1,y_2)$ in $\mathbb{R}^2$.
The same integration over the coordinates $(x_1,x_2)$ performed on $R^{\hbar,\theta}_{r^{\hbar,\theta}_t(r_0)}(r)$ 
in \eqref{wigt} yields the following reduced Wigner function
at time $t\neq 0$:
\begin{eqnarray}\nonumber
R^{\hbar,\theta}_{x_0,y_0}(y,t) &:=& \int_{\mathbb{R}^2}{\rm d}x_1{\rm d}x_2 R^{\hbar,\theta}_{r^{\hbar,\theta}_{-t}(r_0)}(x_1,x_2,y_1,y_2)\\\label{wigt1}
 &=& \frac{1}{\pi m^2\omega^2\theta} \,{\rm e}^{-\frac{1}{m^2\omega^2\theta}\left((y_1 -y^{\hbar,\theta}_{10}(-t))^2 + 
(y_2-y^{\hbar,\theta}_{20}(-t))^2\right)}\ ,
\end{eqnarray}
where $x_0=(x_{10},x_{20})$.

Notice that, because of the time-evolution, through $y^{\hbar,\theta}_{10}(-t)$ and $y^{\hbar,\theta}_{20}(-t)$, 
the reduced Wigner function at time $t$ is not a function of the momentum initial condition $y_0$, only, but also 
of the coordinate initial condition $x_0$.
Such a dependence does not disappear by letting $\hbar\to0$; indeed, by using the time-evolution matrix $A^{0,\theta}_t = 
\lim_{\hbar\to0}A^{\hbar,\theta}_t$ in \eqref{A24}, one gets that the initial condition
$R^{0,\theta}_{y_0}(y)=\lim_{\hbar\to0}R^{\hbar,\theta}_{y_0}(y)$ goes into
\begin{eqnarray}\nonumber
R^{0,\theta}_{x_{10},y_0}(y,t) &=& \lim_{\hbar\to0}R^{\hbar,\theta}_{x_0,y_0}(y,t)\\\label{wigt2}
 &=& \frac{1}{\pi m^2\omega^2\theta}\,
{\rm e}^{-\frac{1}{m^2\omega^2\theta}\left(\left(y_1 -y_{10}(\cos\omega t)+x_{10}\sin(\omega t)\right)^2 + (y_2-y_{20})^2\right)}.
\end{eqnarray}
In order to get a meaningful dynamical map connecting the Wigner function at time $t=0$, $R^{0,\theta}_{y_0}(y)$, to a function of 
time $t$, momentum $y\in\mathbb{R}^2$ and initial momentum $y_0\in\mathbb{R}^2$, one has to eliminate the dependence on the parameter 
$x_{10}$. This can only be done by a further integration over $y_1$ which leaves us with 
$$
R^{0,\theta}_{y_{20}}(y_2)=\frac{1}{\sqrt{\pi\theta} m\omega}\,
{\rm e}^{-\frac{1}{m^2\omega^2\theta}\left(y_2-y_{20}\right)^2}\ ;
$$ 
namely, with a probability distribution over $\mathbb{R}$ only, with no dynamics anymore, exactly as in \eqref{nodyn}.
Differently from the previous dimensionality reduction from $\mathbb{R}^4$ to $\mathbb{R}^2$ which was entirely due to 
kinematical reasons, the further one from $\mathbb{R}^4$ to $\mathbb{R}$ is a consequence of the dynamics that mixes position and momentum coordinates. 

\section{Concluding remarks}\label{sec5}

This paper provides an interpretation in terms of phase-space (semi-classical) localisation properties 
of the fact that, when dealing with two quantum oscillators 
(parameter $\hbar$) on a noncommutative configuration-space (parameter $\theta$), 
one finds that $\lim_{\theta\to 0}\lim_{\hbar\to 0}\neq\lim_{\hbar\to 0}\lim_{\theta\to0}$.
Namely, that the limit corresponding to the removal of 
noncommutativity from the configuration space ($\theta\to0$) does not commute with the usual classical 
limit that removes position and momentum noncommutativity ($\hbar\to0$).

We have shown that the nonexchangeability of the two limits is a purely kinematical effect related to the different 
phase-space localisation properties of coherent states when $\theta\to0$ and $\hbar\neq0$ with respect to when $\hbar\to0$ and $\theta\neq 0$.
In the first case, the limit yields the standard coherent state Wigner function of two quantum harmonic oscillators, while, in the second case, 
a kinematical description is only possible by going to a reduced Wigner function integrated over the $2$-dimensional position space.   
The necessity of reducing the domain of definition of the Wigner function from $\mathbb{R}^4$ to $\mathbb{R}^2$ 
becomes more dramatic when the free noncommutative quantum oscillator dynamics is taken into account. 
Since initial position and momentum operators are mixed in the course of time,  if 
operator noncommutativity is removed before configuration-space noncommutativity, a Wigner distribution can only survive 
if a further integration is performed on the first momentum-like coordinate. However, this integration provides 
a well-defined distribution over $\mathbb{R}$ at the price of completely eliminating any time-dependence.

\section{Appendix}

We list here the behaviour of the various $\hbar,\theta$-dependent quantities when $\theta\to0$ and $\hbar\to 0$, respectively.
\begin{itemize}
\item
$\theta\to0$:
\begin{eqnarray}
\label{A1}
&&
\lambda^{\hbar,\theta}_\pm=\frac{1}{2}\left(
m\omega\sqrt{4\hbar^2+m^2\omega^2\theta^2}\pm m^2\omega^2\theta
\right)\to m\omega\hbar\\
\label{A2}
&&\lambda^{\hbar,\theta}_++\lambda^{\hbar,\theta}_-=
m\omega\sqrt{4\hbar^2+m^2\omega^2\theta^2}\to 2m\omega\hbar\\
\label{A3}
&&
\mu^{\hbar,\theta}=\frac{\lambda^{\hbar,\theta}}{m\omega}
\to \hbar\\
\label{A4}
&&
K^{\hbar,\theta}_\pm=\lambda^{\hbar,\theta}_\pm\left(
4\pm\frac{2\theta\lambda_\pm^{\hbar,\theta}}{\hbar^2}\right)\to 4m\omega\hbar
\end{eqnarray}
\item
$\hbar\to0$:
\begin{eqnarray}
\label{A5}
&&
\lambda^{\hbar,\theta}_+=\frac{1}{2}\left(
m\omega\sqrt{4\hbar^2+m^2\omega^2\theta^2}+ m^2\omega^2\theta
\right)\to m^2\omega^2\theta\\
\label{A6}
&&
\lambda^{\hbar,\theta}_-=\frac{1}{2}\left(
m\omega\sqrt{4\hbar^2+m^2\omega^2\theta^2}- m^2\omega^2\theta
\right)\to \hbar^2/\theta\\
\label{A7}
&&
\lambda^{\hbar,\theta}_++\lambda^{\hbar,\theta}_-=
m\omega\sqrt{4\hbar^2+m^2\omega^2\theta^2}\to m^2\omega^2\theta\\
\label{A8}
&&
\mu^{\hbar,\theta}=\frac{\lambda^{\hbar,\theta}}{m\omega}
\to m\omega\theta\\
\label{A9}
&&
K^{\hbar,\theta}_+=\lambda^{\hbar,\theta}_+\left(
4+\frac{2\theta\lambda_+^{\hbar,\theta}}{\hbar^2}\right)\to \frac{2m^4\omega^4\theta^3}{\hbar^2}\\
\label{A10}
&&
K^{\hbar,\theta}_-=\lambda^{\hbar,\theta}_-\left(
4-\frac{2\theta\lambda_-^{\hbar,\theta}}{\hbar^2}\right)\to \frac{2\hbar^2}{\theta}\ .
\end{eqnarray}
\end{itemize}
From these it follows that the matrix
$$
\hat{J}^{\hbar,\theta} = \frac{1}{2\mu_{\hbar,\theta}(\lambda^{\hbar,\theta}_{+} +\lambda^{\hbar,\theta}_{-})}
\begin{pmatrix}
 \lambda^{\hbar,\theta}_{-}\sqrt{K^{\hbar,\theta}_+} & 0 & 0 & -\hbar\sqrt{K^{\hbar,\theta}_+}\cr
 -\lambda^{\hbar,\theta}_+\sqrt{K^{\hbar,\theta}_{-}} & 0 & 0 & -\hbar\sqrt{K^{\hbar,\theta}_{-}}\cr
 0 & \lambda^{\hbar,\theta}_{-}\sqrt{K^{\hbar,\theta}_+} & \hbar\sqrt{K^{\hbar,\theta}_+} & 0\cr
 0 & \lambda^{\hbar,\theta}_{+}\sqrt{K^{\hbar,\theta}_{-}} & -\hbar\sqrt{K^{\hbar,\theta}_{-}} & 0
\end{pmatrix}
$$
has the following limits
\begin{eqnarray}
\label{A11}
\hat{J}^{\hbar,0}&:=&\lim_{\theta\to0}\hat{J}^{\hbar,\theta}=\frac{1}{2\sqrt{\hbar}}\begin{pmatrix}
\sqrt{m\omega}& 0 & 0 & -\frac{1}{\sqrt{m\omega}}\cr
-\sqrt{m\omega}& 0 & 0 & -\frac{1}{\sqrt{m\omega}}\cr
 0 & \sqrt{m\omega}& \frac{1}{\sqrt{m\omega}}& 0\cr
 0 & \sqrt{m\omega} & -\frac{1}{\sqrt{m\omega}}& 0
\end{pmatrix}\\
\nonumber
\\
\nonumber
\label{A12}
\hat{J}^{0,\theta}&:=&\lim_{\hbar\to0}\hat{J}^{\hbar,\theta}=\frac{1}{m\omega\sqrt{2\theta}}\begin{pmatrix}
0 & 0 & 0 & -1\cr
0 & 0 & 0 & 0\cr
0 & 0 & 1 & 0\cr
0 & 0 & 0 & 0
\end{pmatrix}\ .
\end{eqnarray}

Concerning the behaviour of the functions appearing in \eqref{f12}--\eqref{g12}, we have
\begin{itemize}
\item
$\theta\to0$:
\begin{eqnarray}
\label{A13}
&&
\gamma^{\hbar,\theta}_\pm=\frac{1}{2}\Big(1 \pm\frac{m\omega\theta}
{\sqrt{4\hbar^2 + 2m^2\omega^2\theta^2}}\Big)\to \frac{1}{2}\\\nonumber
\label{A14}
&&
f^{\hbar,\theta}(x,y)=\frac{\mu^{\hbar,\theta}\sqrt[4]{4\hbar^2 + m^2\omega^2\theta^2}}{2\sqrt{m\omega}\hbar}
\left(\frac{x}{\sqrt{\gamma^{\hbar,\theta}_+}}+\frac{y}{\sqrt{\gamma^{\hbar,\theta}_{-}}}\right)
\to \sqrt{\frac{\hbar}{m\omega}}(x+y)=:f^{\hbar,0}(x,y)\\
\\\nonumber
&&
g^{\hbar,\theta}(x,y)=\frac{\mu^{\hbar,\theta}\sqrt{m\omega}}{\sqrt[4]{4\hbar^2 + m^2\omega^2
\theta^2}}\left(\frac{x}{\sqrt{\gamma^{\hbar,\theta}_+}}-
\frac{y}{\sqrt{\gamma^{\hbar,\theta}_{-}}}\right)\to
\sqrt{m\omega\hbar}(x-y)=:g^{\hbar,0}(x,y).\\\label{A15}
\end{eqnarray}
\item
$\hbar\to0$:
\begin{eqnarray}
\label{A16}
&&
\hskip-1cm
\gamma^{\hbar,\theta}_\pm=\frac{1}{2}\Big(1 \pm\frac{m\omega\theta}
{\sqrt{4\hbar^2 + 2m^2\omega^2\theta^2}}\Big)\to \frac{1}{2}\left(1\pm\frac{1}{\sqrt{2}}\right)=:\gamma_\pm\\
\label{A17}
&&
\hskip-1cm
f^{\hbar,\theta}(x,y)=\frac{\mu^{\hbar,\theta}\sqrt[4]{4\hbar^2 + m^2\omega^2\theta^2}}{2\sqrt{m\omega}\hbar}
\left(\frac{x}{\sqrt{\gamma^{\hbar,\theta}_+}}+\frac{y}{\sqrt{\gamma^{\hbar,\theta}_{-}}}\right)\to
\frac{m\omega}{2}\frac{\theta^{3/2}}{\hbar}\left(\frac{x}{\sqrt{\gamma_+}}+\frac{y}{\sqrt{\gamma_-}}\right)\\\nonumber
\label{A18}
&&
\hskip-1cm
g^{\hbar,\theta}(x,y)=\frac{\mu^{\hbar,\theta}\sqrt{m\omega}}{\sqrt[4]{4\hbar^2 + m^2\omega^2
\theta^2}}\left(\frac{x}{\sqrt{\gamma^{\hbar,\theta}_+}}-
\frac{y}{\sqrt{\gamma^{\hbar,\theta}_{-}}}\right)\to
m\omega\sqrt{\frac{\theta}{2}}\left(\frac{x}{\sqrt{\gamma_+}}-\frac{y}{\sqrt{\gamma_-}}\right)=:g^{0,\theta}(x,y).\\ 
\end{eqnarray}
\end{itemize}

Regarding the time-evolution in the various limits, one has:
\begin{itemize}
\item
$\theta\to0$:
\begin{eqnarray}
\label{A20}
&&\omega_\pm^{\hbar,\theta}=\frac{\lambda_\pm^{\hbar,\theta}}{m\mu^{\hbar,\theta}}\to \omega\\
\label{A21}
&&
A^{\hbar,\theta}_t \to
A^{\hbar,0}_t = \left(
 \begin{array}{cccc}
  \cos\omega t & 0 & -\sin\omega t & 0\\
  0 & \cos\omega t & 0 & -\sin\omega t\\
  \sin\omega t & 0 & \cos\omega t & 0\\
  0 & \sin\omega t & 0& \cos\omega t
\end{array}
\right)
\end{eqnarray}
\item
$\hbar\to0$:
\begin{eqnarray}
\label{A22}
&&
\omega_+^{\hbar,\theta}=\frac{\lambda_+^{\hbar,\theta}}{m\mu^{\hbar,\theta}}\to \omega\\
\label{A23}
&&
\omega_-^{\hbar,\theta}=\frac{\lambda_-^{\hbar,\theta}}{m\mu^{\hbar,\theta}}\to \frac{\hbar^2}{m^2\omega^2\theta^2}\\
\label{A24}
&&
A^{\hbar,\theta}_t \to A^{0,\theta}_t=\left(
 \begin{array}{cccc}
  \cos\omega t & 0 & -\sin\omega t & 0\\
  0 & 1 & 0 & 0\\
  \sin\omega t & 0 & \cos\omega t & 0\\
  0 & 0 & 0& 1
\end{array}
\right)
\end{eqnarray}
\end{itemize}

{\bf Acknowlegment}:

LG is supported by {\it The Abdus Salam International Centre for Theoretical Physics} (ICTP).

\end{document}